\documentclass[sigconf]{acmart}
\usepackage{tabularx}

\acmConference[TAKEDOWN '26]{ACM Workshop on Technical Analysis and Knowledge Exchange on Disrupting Online Criminal Networks}{November 15--19, 2026}{The Hague, The Netherlands}
\acmYear{2026}
\copyrightyear{2026}
\acmDOI{10.1145/nnnnnnn.nnnnnnn}
\acmISBN{978-1-4503-XXXX-X/26/11}

\usepackage{booktabs}
\usepackage{array}
\usepackage{enumitem}
\usepackage{xcolor}
\usepackage{listings}
\usepackage{todonotes}

\usepackage{tcolorbox}
\usepackage{listings}

\makeatletter
\renewcommand{\ACM@linecountL}{}
\renewcommand{\ACM@linecountR}{}
\makeatother

\begin{document}

\title{A Bulletproof Business? Towards Detecting Infrastructure-as-a-Service Offerings on Telegram}

\author{Roy Ricaldi}
\affiliation{%
  \institution{Eindhoven University of Technology}
  \city{Eindhoven}
  \country{The Netherlands}
}
\email{r.j.ricaldi.saavedra@tue.nl}

\author{Kristiyan Kyurkchiev}
\affiliation{%
  \institution{Eindhoven University of Technology}
  \city{Eindhoven}
  \country{The Netherlands}
}
\email{k.kyurkchiev@student.tue.nl}

\author{Irdin Pekaric}
\affiliation{%
  \institution{University of Liechtenstein}
  \city{Vaduz}
  \country{Liechtenstein}
}
\email{irdin.pekaric@uni.li}

\hypersetup{
  pdfauthor={Roy Ricaldi, Kristiyan Kyurkchiev, Irdin Pekaric}
}

\renewcommand{\shortauthors}{Ricaldi et al.}

\begin{abstract}
Cybercriminal operations increasingly rely on reusable digital infrastructure -- such as hosting, proxies and VPNs --- rented through Cybercrime-as-a-Service markets and advertised on platforms including Telegram. We develop a taxonomy for labeling Telegram messages that advertise cybercriminal Infrastructure-as-a-Service organized into compute, network and communication dimensions including three trust attributes: Bulletproof, Payment Security and Transparency. We compare keyword-based, TF-IDF and prompt-based LLM classifiers and apply the selected TF-IDF classifier to 1,116,071 messages from 167 cybercriminal communities. The classifier assigns at least one infrastructure category to 207,244 messages (18.57\%) across 113 communities. Advertising is highly concentrated: one community accounts for 50.3\% of classified infrastructure messages, while Bulletproof claims appear in 37.66\%. These results identify high-volume communities and advertising identities as priorities for monitoring, investigation and potential disruption.

\end{abstract}



\maketitle

\section{Introduction}
\label{sec:intro}

Cybercrime does not run on malicious software alone. For example, phishing depends --- beside malicious software --- on reusable digital infrastructure such as hosting, virtual private servers, remote desktop access, proxies, VPNs and mail-sending systems. These resources provide computing capacity, connectivity, anonymity and communication capabilities that can be rented through Cybercrime-as-a-Service markets~\cite{akyazi2021caas,van2018plug}. Because the same infrastructure may support multiple customers and criminal operations, its disruption can have broader impacts~\cite{collier2021boring, georgoulias2023botnet, vu2025assessing,ricaldi2026topical}.

Telegram has become an important venue for advertising such services. Its pseudonymous accounts, large communities, bot support and limited moderation attract actors involved in malware distribution, fraud, stolen-data sales and other illicit activities~\cite{schroer2025dark}. However, Telegram messages are less structured than conventional marketplace listings. Advertisements, conversations, forwarded posts and trust claims appear in the same message streams, while vendors bundle several services in a single offer.

Prior work has developed detailed Cybercrime-as-a-Service taxonomies using forum data~\cite{akyazi2021caas}, examined infrastructure as a target for cybercrime disruption~\cite{moura2024characterizing, sokoto2024guardians, bouwman2025can} or analyzed trust signals used by Telegram vendors~\cite{ricaldi2025trust}. However, prior approaches do not distinguish between the specific infrastructure services advertised on Telegram. They also overlook bulletproof claims via which vendors market resistance to abuse complaints, suspension or takedown.

We address this gap by developing a Telegram-specific taxonomy and detection pipeline for cybercriminal Infrastructure-as-a-Service offerings. 
Our results show how large-scale message classification can support the prioritization of communities and service categories for further investigation and potential takedown efforts.

\noindent \textsc{\textbf{Contributions.}}
We make the following contributions:
\begin{itemize}[leftmargin=*,noitemsep,topsep=0pt]

\item \textbf{A Telegram-specific taxonomy and detection pipeline for cybercriminal digital infrastructure} (RQ1): We organize six service categories into three capability-based families: \textit{Compute}, \textit{Network} and \textit{Communication Infrastructure}. In addition, we incorporate three trust attributes, including \textit{Bulletproof}, which captures vendors’ claims of resilience to abuse complaints. 
\item \textbf{A large-scale prevalence measurement with takedown implications and data access to support research} (RQ2): We apply the best-performing classifier to 1,116,071 Telegram messages across 167 cybercriminal communities, where we show that $\approx 19\%$ of messages advertise at least one infrastructure category and that advertising activity is highly concentrated -- a single group accounts for over half of all classified infrastructure messages -- providing a data-driven basis for prioritizing monitoring and takedown efforts.
\item \textbf{An empirically-grounded roadmap for extending the taxonomy}: We identify specific infrastructure families in the literature and during annotation, such as certificate, storage and command-and-control infrastructure that current data cannot yet support, but that future annotation efforts with a larger dataset should target and take into consideration.

\end{itemize}
\vspace{2mm}
We release our resources to support research transparency~\cite{pekaric2025transparency} and software sustainability~\cite{pekaric2026fitness}:
\url{https://github.com/irdin-pekaric/TAKEDOWN2026}.

\section{Background and Related Work}
\label{sec:related}
\textbf{Cybercrime-as-a-Service.} Cybercrime increasingly depends on specialized, reusable resources that are rented rather than developed by individual offenders --- a model termed Cybercrime-as-a-Service (CaaS) that lowers the barrier to participation~\cite{akyazi2021caas} and creates shared dependencies across otherwise unrelated actors~\cite{huang2015framing}. Shared services such as hosting, remote access, proxies, VPNs and mail-sending are reused across operations, which makes their providers potentially valuable targets for disruption~\cite{collier2021boring,noroozian2019platforms,kropotov2020hacker}. In this regard, Akyazi et al.~\cite{akyazi2021caas} develop a taxonomy of CaaS advertisements and compare supervised methods for classifying them, but their analysis is based on forum data and not on Telegram data. 

\textbf{Cybercrime on Telegram.}
Telegram’s low barriers to entry, large channels and instant messaging have made it a prominent venue for cybercrime advertising~\cite{allodi2024dmitry,ricaldi2026telehunt}. Unlike conventional forums, however, its chronological and unstructured message streams combine advertisements, conversations, forwarded posts and other content, making individual offerings more difficult to identify~\cite{hughes2024art}. TeleHunt classifies communities into cybercriminal market segments, including Digital Infrastructure~\cite{ricaldi2026telehunt}, and DarkGram classifies channels into activity categories~\cite{roy2025darkgram}; however, neither distinguishes the individual infrastructure services advertised in those categories. Automated classification at finer granularity remains difficult regardless: Hughes et al.~\cite{hughes2024art} identify unreliable ground truth and rapidly-evolving, community-specific terminology as persistent obstacles in cybercrime community research.
These challenges motivate our comparison of multiple detection approaches instead of assuming that a method developed for forums or broader Telegram categories transfers directly to infrastructure advertisements.

\textbf{Trust Signaling in Underground Markets.} On forums, Hughes et al.~\cite{hughes2023argot} show that argot itself functions as a trust signal. They apply signalling theory, where they find argot use is negatively correlated with reputation and helps newcomers overcome the ``cold start'' problem of building trust without an existing feedback history. Closest to our work, Ricaldi et al.~\cite{ricaldi2025trust} identify 11 trust signals across six segments of Telegram’s cybercrime economy, covering reputation, payment security, moderation and transparency and scale their framework using a fine-tuned LLM classifier. They find that the Infrastructure segment has the highest average exposure to trust signals, driven mainly by payment security and transparency. Their framework, however, does not capture ``bulletproof'' claims through which infrastructure vendors advertise resistance to abuse complaints, service suspension and takedowns.

\textbf{Research Gap.} Prior work either classifies Telegram communities broadly~\cite{ricaldi2026telehunt,roy2025darkgram}, classifies CaaS offerings in more detail but on forum data~\cite{akyazi2021caas}, examines shared infrastructure and disruption without a Telegram-specific measurement method~\cite{collier2021boring,noroozian2019platforms} or maps trust signaling across Telegram without modeling infrastructure-specific claims such as bulletproofness~\cite{ricaldi2025trust}. No prior work jointly identifies specific infrastructure services advertised on Telegram, detects them at scale and examines how their distribution and associated trust signals can inform disruption priorities. Hence, our work is guided by the following RQs:

\begin{itemize}[leftmargin=*,noitemsep,topsep=0pt]
    \item\textbf{RQ1:} What types of Infrastructure-as-a-Service are advertised on Telegram, and how can they be reliably categorized and automatically detected?
    \item\textbf{RQ2:} How are these Infrastructure-as-a-Service offerings and trust signals distributed and concentrated across Telegram vendors and communities, and what do these patterns imply for disruption prioritization?
\end{itemize}

\section{Methodology}
\label{sec:methodology}
We define the study scope and dataset (§\ref{sec:scope_data}), then develop and validate the infrastructure and trust-signal taxonomy (§\ref{sec:groundtruth}). Next, we compare, implement and validate automated multi-label detection approaches (§\ref{sec:labeling}), while considering ethical considerations (§\ref{ssec:ethics}).

\subsection{Scope and Data}
\label{sec:scope_data}
We define digital infrastructure as \textit{reusable technical resources that provide computing, remote access, routing, anonymity or message-delivery capabilities for cybercriminal operations}. The data set we use is drawn from \cite{ricaldi2025trust} and it contains 1,116,071 messages from 167 Telegram groups and channels, spanning January 2023--March 2025 with per-message metadata: the originating Telegram group or channel and sending user, and a market segment label for the group\footnote{Ricaldi et al. identify six market segments in the Telegram cybercrime economy: Cyberattacks, Fraud, Digital Infrastructure, Personal Data, Piracy and Tutorials.}. The source study \cite{ricaldi2025trust} seeded 30 communities from three cybercrime forums and Telegram keyword searches, then expanded the collection by following group and channel handles or links mentioned in messages, and collected their messages using an automated tool. This process includes private Telegram communities which do not have public handles and can only be joined by clicking on a private link circulated by the administrators. We use “community” to refer collectively to a Telegram group or channel. The IaaS taxonomy is developed using a \textbf{subset} of the dataset, which includes only communities focused on digital infrastructure. This subset contains 209,803 messages from 11 Telegram communities and 5,030 unique users. 
These messages form the corpus used to iteratively develop and validate the proposed taxonomy. Details on how Ricaldi et al. collected and curated this dataset can be found in App. \ref{app:trust_signals_dataset}.

\subsection{Taxonomy Development}
\label{sec:groundtruth}
\textbf{Infrastructure Taxonomy.} Following practices from prior work on cybersecurity data labeling~\cite{braun2024understanding}, four independent coders annotate the types of offerings in the data using a stratified\footnote{The sample is stratified by community to ensure representation from all 11 communities, preventing domination by the largest communities.} sample of 500 messages from the subset. Coders derive an initial taxonomy of 12 families and 48 subcategories (App.~\ref{app:initial-taxonomy}). They scope the taxonomy down to categories substantively relevant to their definition of digital infrastructure. Coders organize the remaining categories by the operational capability that the infrastructure provides to a buyer. In parallel, trust signals in the message offerings are annotated to understand how vendors support the trade of digital infrastructure on Telegram. Lastly, 300 messages are coded using the working taxonomy of infrastructure types and trust signals to verify its stability, with no new categories arising in the process. 

\noindent \textbf{Taxonomy Validation.} After stabilizing the taxonomy, two coders independently annotated (multi-labeled) a stratified\footnote{Messages were selected via category-relevant keywords of each infrastructure type.} sample of 250 messages from the subset focusing on infrastructure categories and three trust signals. Cohen's kappa was high for five of six infrastructure
categories ($\kappa = 0.74$–$0.98$), with Proxy the exception at
moderate agreement ($\kappa = 0.58$), and for all three trust
attributes ($\kappa = 0.64$–$0.90$); overall agreement was strong
(macro $\kappa = 0.81$, micro $\kappa = 0.86$). Full results appear in
App. C.1 (Tab.~\ref{tab:iaa}).

\subsection{Model Implementation}
\label{sec:labeling}

We formulate infrastructure detection as multi-label text classification. The 250-message taxonomy-validation sample was adjudicated by two annotators and supplemented with 11 jointly annotated proxy messages, yielding 261 labeled messages. We compared a keyword dictionary and TF-IDF with logistic regression using a fixed 75\%/25\% training/test split, fitting both approaches only on the training partition. We also explored prompt-based classification using locally deployable, open-weight LLMs, motivated by prior work on NLP-based detection and LLMs in cybersecurity~\cite{pekaric2026llms, ave2026bot}. Tab.~\ref{tab:classifier-comparison-main} summarizes the lightweight classifier comparison. TF-IDF achieved higher micro-F1, precision and exact-match accuracy, whereas the keyword approach achieved higher macro-F1 and recall. We selected TF-IDF to prioritize aggregate precision and micro-F1.

\begin{table}[t]
    \centering
    \caption{Infrastructure classification on the held-out test split.
    Precision and recall are micro-averaged; Exact denotes exact-match accuracy.}
    \label{tab:classifier-comparison-main}
    \small
    \setlength{\tabcolsep}{3pt}
    \begin{tabular}{lccccc}
        \toprule
        Approach & Micro-F1 & Macro-F1 & Prec. & Rec. & Exact \\
        \midrule
        Keyword     & 0.847 & 0.759 & 0.746 & 0.979 & 0.652 \\
        TF-IDF + LR & 0.898 & 0.716 & 0.880 & 0.917 & 0.803 \\
        \bottomrule
    \end{tabular}
\end{table}


\subsubsection{TF-IDF with Logistic Regression Model} For the selected TF-IDF approach, messages were normalized and represented using word-level unigrams and bigrams. The vectorizer retained at most 20,000 features, excluded terms occurring in more than 98\% of the training messages, and applied sublinear term-frequency scaling. One binary logistic-regression classifier was trained per label using a one-vs-rest strategy, balanced class weights, the \texttt{liblinear} solver, and \(C=2.0\). Separate models for infrastructure categories and trust attributes were trained using a human-annotated set of 261 messages from a sample stratified by community of the 11 Digital Infrastructure Telegram communities in our subset (§\ref{sec:scope_data}). Labels were assigned independently by the model using a probability threshold of 0.5\footnote{A threshold of 0.5 was used for all labels, following the standard decision rule for logistic regression. We did not tune thresholds to avoid introducing additional model-specific optimization.}, allowing a message to receive zero, one, or multiple labels. This model was applied to all 1,116,071 messages. The implementation details may be found in App. \ref{app:automation_methods}.

\subsubsection{Model Validation.} We independently checked infrastructure predictions on 200 additional messages: 100 predicted-positive messages from communities outside the training domain, 50 predicted-positive messages from the training domain, and 50 predicted-negative messages. Precision for detecting any infrastructure offering was 78\% outside the training domain and 84\% within it. Human annotation identified infrastructure in 5 of the 50 predicted-negative messages; this is the observed miss share among sampled negative predictions, not an estimate of population recall. A separate 100-message sample, stratified by predicted trust attributes, yielded pooled Cohen's $\kappa = 0.615$. Payment Security was less reliable (precision $= 0.343$) than Bulletproof ($0.769$) and Transparency ($0.833$). These checks support exploratory analysis while identifying uncertainty in category- and attribute-specific estimates.

\subsection{Ethical Considerations} 
\label{ssec:ethics}

We reuse the dataset from~\cite{ricaldi2025trust} under an agreement permitting scientific research, without interacting with community members or collecting additional Telegram data. Our unit of analysis is the advertised \emph{infrastructure}, not personal data: the taxonomy and codebook (see App.~\ref{app:codebook}) target technical categories such as hosting and tunnels, and we do not analyze or report on the content of stolen data, victims or other non-infrastructure offers that may appear in the same channels. We publish only aggregate, category- and group-level statistics and anonymize advertising identities. We assess the risk of operational uplift as limited because the released taxonomy and keywords describe terminology already used openly in vendor advertisements. Their disclosure primarily supports replication and independent evaluation of the measurement method~\cite{menlo2012}.


\section{Results}
\label{sec:results}

We present the resulting infrastructure and trust-signal taxonomy (§\ref{sec:results_taxonomy}), then examine the prevalence and concentration of infrastructure advertising across messages, advertising identities and communities (§\ref{sec:results-prevalence}–§\ref{sec:results-concentration}). Finally, we analyze how trust signaling varies across infrastructure categories and how advertising volume and specialization reveal different priorities for further investigation and disruption (§\ref{sec:results-trust}–§\ref{sec:results-disruption}).

\subsection{A Taxonomy of Infrastructure-as-a-Service Offerings on Telegram (RQ1)} \label{sec:results_taxonomy}

Tab.~\ref{tab:taxonomy} presents six infrastructure categories organized into Compute, Network and Communication Infrastructure, alongside three non-exclusive trust attributes. Infrastructure labels identify the advertised capability, whereas trust attributes describe claims intended to establish credibility. Messages may receive multiple labels, but Hosting is assigned alongside a specific product only when general hosting is explicitly offered. Communication Infrastructure currently contains Mail infrastructure; additional communication services can be incorporated when sufficient examples support reliable annotation. The full codebook in App.~\ref{app:codebook} provides operational definitions and boundary rules.

\begin{table}[t]
\centering
\caption{Taxonomy of infrastructure and trust.}
\label{tab:taxonomy}
\small
\textbf{(a) Infrastructure categories}
\vspace{0.4em}
\begin{tabular}{@{}p{0.5cm}p{1cm}p{6cm}@{}}
\toprule
\textbf{Code} & \textbf{Category} & \textbf{Definition} \\
\midrule
\multicolumn{3}{@{}l}{\emph{Compute Infrastructure}}\\
1.1 & Hosting & General-purpose hosting, cloud, or dedicated server infrastructure (not specific like below).\\
1.2 & VPS & Virtual private servers or virtual machines, sold as a distinct product.\\
1.3 & RDP & Access to a Windows machine via Remote Desktop Protocol, sold as a ready-to-use login.\\
\addlinespace
\multicolumn{3}{@{}l}{\emph{Network Infrastructure}}\\
2.1 & Proxy services & IP-routing infrastructure that masks, rotates, or geographically targets a connection's origin, per request.\\
2.2 & VPN and tunnels & Infrastructure routing a user's traffic through an intermediary connection for a session.\\
\addlinespace
\multicolumn{3}{@{}l}{\emph{Communication Infrastructure}}\\
3.1 & Mail infras. & Infrastructure for sending, relaying, or managing outbound email at volume.\\
\bottomrule
\end{tabular}
\vspace{0.8em}
\textbf{(b) Trust attributes}
\vspace{0.4em}
\begin{tabular}{@{}p{2cm}p{6cm}@{}}
\toprule
\textbf{Attribute} & \textbf{Definition} \\
\midrule
Bulletproof & A claim of resilience to abuse complaints, law-enforcement action, or takedowns.\\
Payment security & A claim of reduced transaction risk, via escrow, crypto payment, or digital wallets.\\
Transparency & A claim of service accountability, via guaranteed support, proof of delivery, or warranties.\\
\bottomrule
\end{tabular}
\end{table}

\begin{table*}[t]
\caption{Predicted infrastructure-category prevalence at message and community level.}
\label{tab:category-prevalence}
\footnotesize
\setlength{\tabcolsep}{8pt}
\begin{tabular}{@{}llrrrrrrrr@{}}
\toprule
Code & Category & Msgs & \%All & \%Infra & Communities & \%AllComms & \%InfraComms & Avg/Comms & Med/Comms \\
\midrule
3.1 & Mail infrastructure & 167,888 & 15.04\% & 81.01\% & 27 & 16.17\% & 23.89\% & 6218.07 & 20.0 \\
1.1 & Hosting services & 123,767 & 11.09\% & 59.72\% & 21 & 12.57\% & 18.58\% & 5893.67 & 13.0 \\
1.3 & RDP & 77,911 & 6.98\% & 37.59\% & 67 & 40.12\% & 59.29\% & 1162.85 & 4.0 \\
1.2 & VPS & 23,699 & 2.12\% & 11.44\% & 37 & 22.16\% & 32.74\% & 640.51 & 3.0 \\
2.1 & Proxy services & 7,412 & 0.66\% & 3.58\% & 42 & 25.15\% & 37.17\% & 176.48 & 3.5 \\
2.2 & VPN and tunnels & 3,963 & 0.36\% & 1.91\% & 83 & 49.70\% & 73.45\% & 47.75 & 5.0 \\
\bottomrule
\end{tabular}
\vspace{1mm}
\begin{minipage}{0.78\textwidth}
\scriptsize
\emph{Note:} ``All groups'' is the share of all Telegram groups containing at least one message from the category. ``Infra. groups'' is the share among groups with at least one predicted infrastructure message. Average and median values are calculated only over groups where the category appears.
\end{minipage}
\end{table*}
\begin{figure}[t]
\centering
\includegraphics[width=\columnwidth]{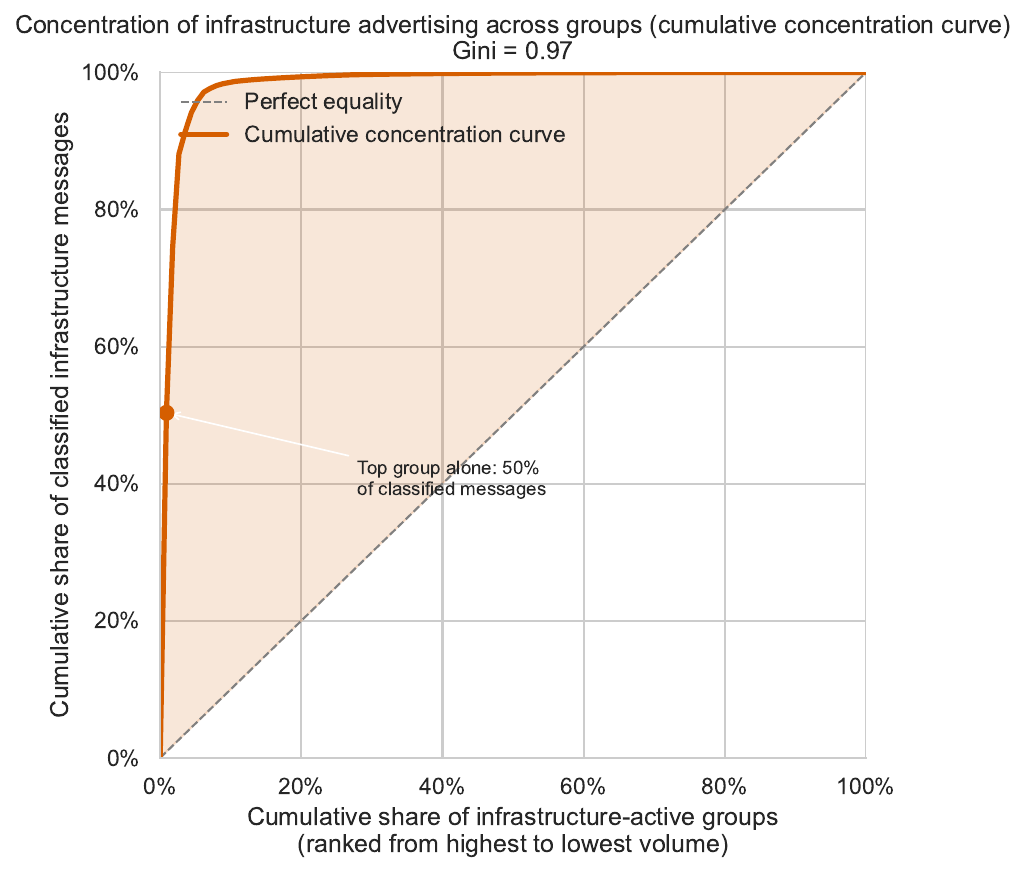}
\caption{Concentration of infrastructure advertising across Telegram groups.}
\Description{A curve showing the concentration of infrastructure advertising across Telegram groups.}
\label{fig:lorenz}
\end{figure}

\subsection{Infrastructure Prevalence (RQ2)}
\label{sec:results-prevalence}

Applying the TF-IDF classifier to all 1,116,071 messages, 207,244 messages (18.57\%) were classified into at least one infrastructure category. Classified messages received 1.95 categories on average, indicating frequent predicted co-occurrence, which may reflect bundled offerings or classification errors.

Tab.~\ref{tab:category-prevalence} reports category prevalence at both message and group level. Message-level prevalence is dominated by Mail infrastructure, which appears in 167{,}888 messages, corresponding to 15.04\% of the full corpus and 81.01\% of classified infrastructure messages. Hosting services follow with 123{,}767 messages, representing 11.09\% of the full corpus and 59.72\% of classified infrastructure messages. RDP appears in 77{,}911 messages, corresponding to 6.98\% and 37.59\%, respectively. VPS, Proxy services and VPN/tunneling are less common, accounting for 2.12\%, 0.66\%, and 0.36\% of the full corpus.

The ordering differs at the community level. VPN/tunneling appears in 83 communities, followed by RDP in 67, Proxy services in 42 and VPS in 37. In contrast, Hosting services and Mail infrastructure appear in only 21 and 27 communities, respectively. High-volume categories are therefore concentrated in relatively few communities, while some lower-volume categories have broader reach\footnote{Community is counted when it has at least one message assigned to a category. Thus, breadth estimates for rare annotated categories should be interpreted cautiously.}. 

\subsection{Advertising-Identity Concentration (RQ2)}\label{sec:results-vendor}
We measure activity by advertising identity: the sending \texttt{user\_id} for 98.09\% of classified messages and the originating channel identity for the remaining 1.91\%. These identities do not necessarily correspond to distinct providers or individuals. Among 207,244 classified messages, we identify 2,010 distinct advertising identities. Concentration at the vendor level is high
(Gini~=~0.884) but less extreme than at the community level
(Gini~=~0.97): the top vendor accounts for 3.43\% of classified
messages, the top~10 for 18.23\%, and the top~50 for 48.16\%.

Among vendors with at least five classified messages (n~=~997),
79.04\% advertise more than one category (mean~=~2.15), while their
dominant category accounts for 64.32\% of messages on average. VPS is
the most common dominant category (31.3\% of vendors) despite ranking
fourth by message volume (Tab.~\ref{tab:category-prevalence}), followed by Mail (25.9\%),
Hosting (19.8\%) and RDP (15.9\%).

Most vendors (87.51\%) advertise in a single community; 12.49\% span
multiple (up to five), including half of the ten highest-volume
vendors, meaning that these sellers would not be fully disrupted by action
against a single community.

Trust signaling scales with vendor volume: Bulletproof use rises from
1.2\% (lowest-volume quartile) to 26.2\% (highest, $\rho = 0.448$),
and Transparency shows a similar gradient (1.4\%--21.5\%,
$\rho = 0.401$). Payment Security is only weakly volume-linked
($\rho = 0.111$), peaking among mid-volume vendors (28.2\%).

\subsection{Community-Level Concentration (RQ2)}
\label{sec:results-concentration}

Infrastructure-related messages appear in 113 of 167 communities (67.66\%). However, the activity is highly uneven. The median infrastructure-active community contains only 11 classified messages, compared with a mean of 1{,}834{.}02, and only five communities contain more than 5{,}000. The highest-volume community accounts for 104{,}320 of 207{,}244 classified infrastructure messages (50.3\%).\footnote{We profile this
community's vendor composition as a case study in App.~\ref{app:case-study}.} As shown in Fig.~\ref{fig:lorenz}, the resulting Gini coefficient of 0.97 indicates extreme concentration, where a small number of communities account for most observed advertising activity. This shows that monitoring or disrupting a limited number of high-volume communities could affect a disproportionate share of observed advertising.



\subsection{Trust Signaling Varies Across Categories}
\label{sec:results-trust}

\begin{figure}[tbp]
\centering
\includegraphics[width=\columnwidth]{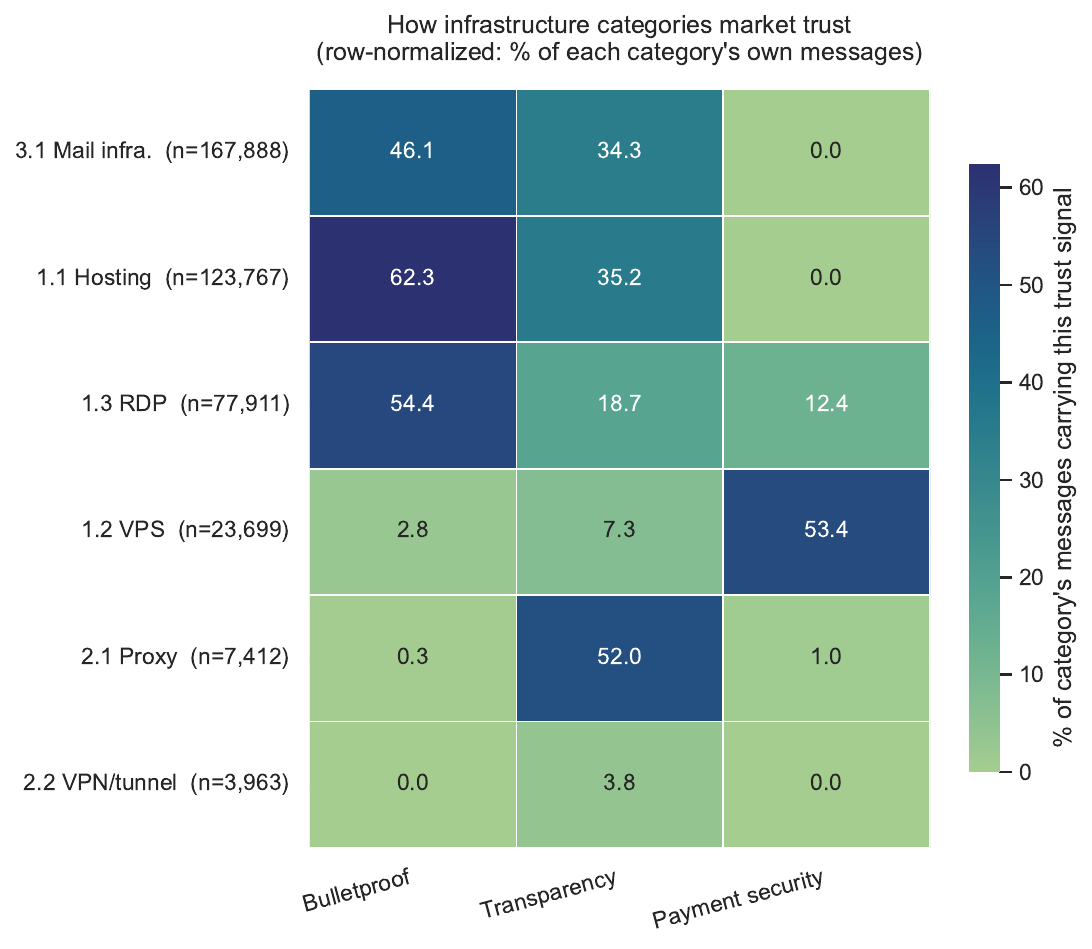}
\caption{Predicted trust attributes within classified infrastructure categories.}
\label{fig:attribute-detail}
\end{figure}

Tab.~\ref{tab:attributes} reports the prevalence of the three trust attributes. Bulletproof is the most common trust signal that appears in 78{,}052 messages, corresponding to 37.66\% of classified infrastructure messages. Transparency appears in 67{,}684 messages or 32.66\%, while Payment Security is less common, appearing in 15{,}310 messages or 7.39\%. Fig.~\ref{fig:attribute-detail} shows that the distribution of trust attributes differs largely across infrastructure categories. Bulletproof is most prevalent in Hosting services (62.3\% of classified messages), followed by RDP at 54.4\% and Mail infrastructure at 46.1\%. These results indicate that claims of resilience to complaints and service disruption are commonly associated with the highest-volume infrastructure categories. Furthermore, Payment Security is most strongly associated with VPS (53.4\% of VPS messages). It also appears in 12.4\% of RDP messages but stays marginal in the other categories. Transparency is most prevalent in Proxy services (52.0\% of classified Proxy messages), followed by Hosting services at 35.2\% and Mail infrastructure at 34.3\%. These predicted associations suggest category-specific trust signaling, although Payment Security estimates require particular caution because of low validation precision.

\begin{table}[t]
\caption{Predicted trust-attribute prevalence at message and group level.}
\label{tab:attributes}
\footnotesize
\setlength{\tabcolsep}{3pt}
\begin{tabular}{@{}lrrrrrr@{}}
\toprule
Attribute & Msgs & \%All & \%Infra & Comms & \%AllC & \%InfC \\
\midrule
Bulletproof & 78,052 & 6.99\% & 37.66\% & 16 & 9.58\% & 14.16\% \\
Transparency & 67,684 & 6.06\% & 32.66\% & 15 & 8.98\% & 13.27\% \\
Payment sec. & 15,310 & 1.37\% & 7.39\% & 21 & 12.57\% & 18.58\% \\
\bottomrule
\end{tabular}
\end{table}

\subsection{Volume and Specialization Indicate Different Disruption Targets}
\label{sec:results-disruption}
\begin{figure}[tbp]
\centering
\includegraphics[width=\columnwidth]{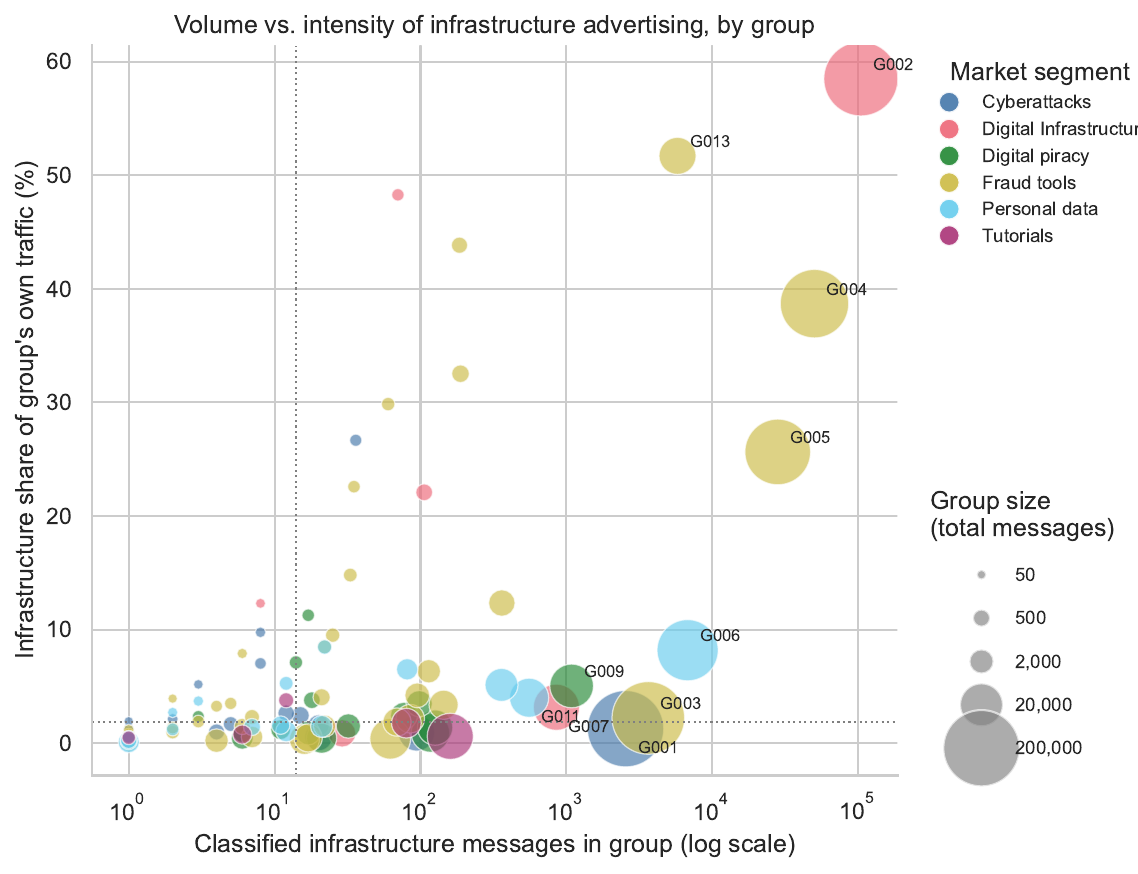}
\caption{Infrastructure advertising volume and specialization across Telegram groups (at least 50 messages).}
\label{fig:intensity}
\end{figure}

Fig.~\ref{fig:intensity} contrasts classified infrastructure-message volume with infrastructure intensity, defined as the share of a community's messages classified as infrastructure. G002 combines high volume with high intensity (approximately 58\%), whereas G006 has substantial volume but only 8\% intensity. G013 exceeds approximately 50\% intensity at lower volume. Volume therefore identifies major advertising venues, while intensity identifies specialized communities. These complementary measures support investigation prioritization; they do not establish the impact of a takedown.

\section{Discussion}
\label{sec:discussion}

We discuss findings on monitoring and disruption prioritization (§\ref{ssec:implications}), limitations (§\ref{ssec:limitations}) and avenues for future research (§\ref{ssec:future}).

\subsection{Implications for Monitoring and Takedown Prioritization}
\label{ssec:implications}

\noindent \textbf{Vendor Diversification.} Cybercriminal infrastructure
advertising on Telegram is driven by a broad vendor base: even the
most active seller accounts for only 3.43\% of classified messages,
and the majority of active vendors (79.04\%) advertise multiple
infrastructure categories rather than a single service. This
suggests that disruption efforts targeting a vendor's primary
offering (e.g., VPS or Mail) may not fully remove the capacity to
supply adjacent services.

\noindent \textbf{Targeting Trust Signals.} Bulletproof claims are concentrated in Hosting services, RDP and Mail infrastructure, which extends prior findings that Infrastructure is Telegram’s most trust-rich market segment~\cite{ricaldi2025trust}. Because bulletproof marketing presents resistance to complaints and takedowns, interventions and public communication could reduce advertising visibility and perceived seller credibility. However, our data captures advertised claims rather than the actual resilience of the underlying infrastructure.

\noindent \textbf{Community-Level Concentration.} Infrastructure advertising is highly concentrated, one community accounts for 50.3\% of all classified infrastructure messages and the distribution has a Gini coefficient of 0.97. Monitoring or disrupting a limited number of high-volume communities could affect a disproportionate share of observed advertising. However, volume alone does not capture specialization, as large general-purpose communities may contain significant infrastructure content. On the other hand, smaller communities may devote their activity to infrastructure advertising.

\noindent \textbf{Volume, Breadth, and Specialization.} Hosting and Mail infrastructure dominate by message volume but appear in relatively few communities, whereas VPN/tunneling and Proxy services have lower volume but broader reach. Communities ranking highly in both volume and infrastructure intensity should be candidates for further investigation, consistent with evidence favoring human-centered integration of AI-driven cyber threat intelligence \cite{karaosman2026security}. Still, removing Telegram communities would primarily disrupt visibility, customer acquisition and trust formation rather than eliminating the underlying infrastructure (advertisers may migrate to other communities or platforms).

Our outputs identify candidates for investigation and do not establish infrastructure ownership. Such prioritization should support evidence-based analyst reasoning, since correct triage decisions need not imply correct explanations~\cite{moosmann2026soc}. Following manual review, advertised domains or endpoints could be examined using passive DNS, TLS certificates and service metadata. Prior work demonstrates how DNS-derived infrastructure features can inform malicious-domain reputation~\cite{antonakakis2010dns}. Such corroboration could reveal technical relationships, but shared hosting or certificates alone would not establish common control.

\subsection{Limitations}
\label{ssec:limitations}

\textbf{Scope and Measurement.} Our results cover only the six infrastructure categories included in the final codebook; excluded, merged or sparsely represented categories are therefore absent from the reported statistics. The classifier was trained with a human annotated sample of messages extracted from 11 Digital Infrastructure communities and then applied to 167 communities across several market segments, which may introduce domain shift. Moreover, our findings reflect classifier predictions rather than complete manual annotation. Estimates for rare categories are less certain, and category co-occurrence is difficult to distinguish from
classifier-induced correlation given overlapping terminology (e.g.,
shared hosting/SMTP vocabulary); further manual validation is therefore needed to determine the extent to which the observed overlap represents genuine service bundling. Community-level results aggregate message predictions; we did not independently evaluate a classifier assigning infrastructure categories directly to communities.



\noindent \textbf{Generalizability.} The dataset contains communities discovered through three cybercrime forums, Telegram keyword searches, and link or handle snowballing. It may not represent closed, vetted, unlinked, or otherwise unobserved communities. The findings may also not generalize to other platforms or later periods, as cybercriminal terminology, service offerings and community structures evolve over time. Finally, our analysis measures advertising activity at the community level and not unique providers or underlying infrastructure ownership (one provider may advertise across several communities, while one community may contain multiple unrelated vendors).


\subsection{Future Work}
\label{ssec:future}

Future work should extend independent validation across communities and time periods, refine weakly performing labels, and add infrastructure categories when sufficient examples become available. Another direction is to investigate whether classified offerings can enrich fragment-based attack models~\cite{pekaric2024streamlining}, connecting advertised infrastructure capabilities to potential attack steps. Within the existing January 2023--March 2025 corpus, monthly analysis could examine changes in trust signals within each infrastructure category while accounting for changing community coverage. Deployment would require periodic annotation of recent messages to assess drift and guide codebook revision and retraining. The present study measures aggregate historical advertising patterns; it does not forecast future offerings or evaluate adaptation to drift.

\section{Conclusion}

This study provides a scalable method for characterizing cybercriminal Infrastructure-as-a-Service advertising on Telegram. We developed a taxonomy of six infrastructure categories and three trust attributes and applied a TF--IDF classification pipeline to 1,116,071 messages from 167 cybercrime-related communities. The pipeline assigned at least one infrastructure category to 207,244 messages (18.57\%) across 113 communities. Advertising was widespread yet highly concentrated: one community accounted for 50.3\% of classified infrastructure messages, while Bulletproof claims appeared in 37.66\% of infrastructure-positive messages. This concentration provides practical starting points for monitoring and investigation. However, the pipeline should support triage rather than be treated as ground truth; establishing offering legitimacy or intervention plans requires manual review and additional evidence.



\section*{Acknowledgments}
We thank Jeroen Oerlemans, Denis Stavinschi, and Jorrit Hofma for their contributions to classifier testing and model development. We are grateful to Tina Marjanov and Jack Hughes of the Cambridge Cybercrime Centre for their assistance with data collection. We also thank Sam Baggen for data management and Jai Wientjes for her input throughout the development of the project; both are members of the Threat Analysis Group at Eindhoven University of Technology. This work was supported in part by Hilti. This publication is part of the INTERSECT project (Grant No.~NWA.1162.18.301), funded by the Dutch Research Council (NWO).

\bibliographystyle{ACM-Reference-Format}
\bibliography{references}

\appendix
\section{Data and Code Availability}
\label{app:availability}

The code, data and documents used for the labeling and testing the tool is available at:
\href{https://github.com/irdin-pekaric/TAKEDOWN2026}{project repository}. 
Because source messages may contain sensitive identifiers, the
message-level dataset is not publicly released; a de-identified
labeled version may be shared for research under the source-data
agreement upon reasonable request.

\subsection{Dataset Provenance}
\label{app:trust_signals_dataset}
The dataset was originally collected by Ricaldi et al.~   \cite{ricaldi2025trust} to map market segments and trust signals in cybercriminal Telegram communities. Their framework identifies six market segments and 11 trust signals. All labels were produced through iterative qualitative coding to saturation drawing various samples of 500 messages from the dataset, then scaled with a DeepSeek-V3 classifier (market-segment F1$=$0.87, $\kappa=0.84$; trust-signal F1$=$0.94, $\kappa=0.76$); see \cite{ricaldi2025trust} for the full annotation and validation procedure. Our subset corresponds to their \emph{Infrastructure} market segment (209{,}803 messages, 11 communities), used for taxonomy development and annotation; the full collection is used only for large-scale classification.

\section{Initial Taxonomy}
\label{app:initial-taxonomy}
Tab.~\ref{tab:initial-taxonomy} lists the full initial taxonomy (12 families, 48 subcategories, plus two catch-all labels) used for exploratory annotation before scoping down to the working taxonomy.

\begin{table}[th]
\caption{Initial cybercrime infrastructure taxonomy used for exploratory annotation.}
\label{tab:initial-taxonomy}
\footnotesize
\begin{tabular}{@{}p{0.30\linewidth}p{0.65\linewidth}@{}}
\toprule
Family & Subcategories \\
\midrule
1. Hosting, Compute, and Storage Infrastructure & 1.1 Bulletproof Hosting; 1.2 General Hosting/VPS/Dedicated Servers; 1.3 Compromised Hosting; 1.4 File, Paste, and Object Storage. \\
2. Network Access, Proxy, and Relay Infrastructure & 2.1 Proxy; 2.2 Remote Access; 2.3 Relay and Anonymization; 2.4 Compromised Network-Device. \\
3. Command, Control, and Botnet Infrastructure & 3.1 Botnet; 3.2 Malware C2; 3.3 Remote Access Malware; 3.4 C2 Resilience. \\
4. Traffic Delivery and Redirection Infrastructure & 4.1 Traffic Redirection; 4.2 Traffic Acquisition; 4.3 Cloaking and Filtering; 4.4 Landing-Page. \\
5. Payload, Malware, and Exploit Delivery Infrastructure & 5.1 Payload Delivery; 5.2 Malware-Spreading; 5.3 Exploit; 5.4 Exploit-Kit/Package. \\
6. Attack-Traffic Infrastructure & 6.1 DDoS/Attack-Traffic; 6.2 Amplification and Reflection; 6.3 Stress-Testing. \\
7. Abuse-Enablement and Anti-Detection Infrastructure & 7.1 Obfuscation/Anti-Detection; 7.2 Security-Checking; 7.3 CAPTCHA-Solving; 7.4 Anti-Fraud Bypass; 7.5 Reputation/Warm-Up. \\
8. Identity, Account, Phone, and Verification Infrastructure & 8.1 Phone/SMS/OTP Verification; 8.2 Account-Creation; 8.3 Identity/KYC-Support; 8.4 Credential and Session. \\
9. Spam, Phishing, and Messaging-Abuse Infrastructure & 9.1 Spam/Email-Abuse; 9.2 Phishing; 9.3 Messaging-App Abuse; 9.4 Notification-Abuse. \\
10. Marketplace, Platform, and Service-Operation Infrastructure & 10.1 Marketplace/Criminal Platform; 10.2 Tool-Pool/Shared Tooling; 10.3 Payment and Escrow; 10.4 Reputation and Access-Control. \\
11. Vulnerability, Scanning, and Security-Testing Infrastructure & 11.1 Vulnerability-Discovery; 11.2 Exploit-Validation; 11.3 Credential/Access-Checking; 11.4 Defensive Security-Checking. \\
12. Social-Platform Telegram-Specific Infrastructure & 12.1 Payment-Bot; 12.2 Content-Access Bot; 12.3 Follow-to-Access Bot; 12.4 Operational Bot. \\
13--14. Other labels & 13. I do not know; 14. Spam/Nonsensical. \\
\bottomrule
\end{tabular}
\end{table}

\subsection{Excluded Categories}
\label{app:exclusion}

We excluded several categories treated as infrastructure in prior CaaS taxonomies: C2/botnet and DDoS/booter infrastructure~\cite{collier2021boring,noroozian2019platforms}, as well as anti-detection/CAPTCHA-solving and account/SMS-verification services~\cite{akyazi2021caas}. Exploratory annotation yielded insufficient examples to support reliable coding, so we leave their inclusion to future work using broader data.


\subsection{Trust Attributes}

Annotators found three trust-related attributes that describe how vendors signal trust: Payment Security, Transparency, and Bulletproof. Payment Security combines the Escrow, Digital Wallets, Cryptocurrency and Automation signals from~\cite{ricaldi2025trust}, while Transparency combines Proof of Delivery, Customer Support, Free Samples and Warranty. They additionally introduce \textbf{Bulletproof} as a new trust signal, absent from their general-purpose framework, to capture marketing that emphasizes resilience to abuse complaints and takedowns.

\section{Codebook}
\label{app:codebook}
Tab.~\ref{tab:codebook} gives the operational definitions used for manual annotation\footnote{All example messages are paraphrased to reduce traceability.} and to build the classifiers, corresponding to the final taxonomy in Tab.~\ref{tab:taxonomy}.

\begin{table*}[t]
\caption{Codebook for the final cybercriminal digital infrastructure taxonomy.}
\label{tab:codebook}
\footnotesize
\begin{tabular}{@{}p{0.03\linewidth}p{0.06\linewidth}p{0.48\linewidth}p{0.19\linewidth}p{0.15\linewidth}@{}}
\toprule
Code & Category & Operational definition & Example message excerpt & Keyword(s) \\
\midrule
\multicolumn{5}{@{}l}{\textit{1. Compute Infrastructure}} \\
1.1 & Hosting & General-purpose hosting, cloud, or dedicated/bare-metal server infrastructure capable of hosting scam pages, panels, bots, or malware. Assign 1.1 together with a more specific category (1.2, 1.3, 2.1, 2.2, or 3.1) only when general hosting is explicitly offered in addition to that product; if the message offers only a VPS, RDP, proxy, VPN, or mail-sending product, assign that specific category alone. & ``cpanel hosting: bulletproof and secure for sensitive projects'' & Hosting, dedicated server, cloud server, cPanel/WHM, bulletproof hosting, offshore hosting \\
1.2 & Virtual private server (VPS) & Virtual private servers or virtual machines sold as a distinct product, typically specifying resources (RAM, vCPU, disk) or OS/root access. Assign alongside 1.1 if general hosting is also explicitly offered, and alongside 1.3 if Remote Desktop access to the VPS is also explicitly offered. & ``wts rdp/vps/servers -- ram 4gb vcpu 2, \$4/month; ram 8gb vcpu 4, \$7/month'' & VPS, VDS, VM, KVM/OpenVZ, vCPU/RAM \\
1.3 & Remote desktop protocol (RDP) & Access to a remote Windows (or Windows-compatible) machine via Remote Desktop Protocol, marketed as a ready-to-use login rather than a general server. Assign regardless of whether the underlying machine is also sold as hosting or a VPS (assign 1.1/1.2 too if explicitly offered). & ``stable 8gb usa rdp and whm available'' & RDP, Windows RDP, Remote Desktop, admin RDP \\
\multicolumn{5}{@{}l}{\textit{2. Network Infrastructure}} \\
2.1 & Proxy services & Intermediary IP-routing infrastructure that masks, rotates, or geographically selects the apparent origin of a connection on a per-request or per-IP basis. Assign 2.2 separately if VPN/tunneling is also explicitly offered. & ``proxy and sop group chat -- join fast, stay tuned!'' & Proxy, residential proxy, rotating IP, IP pool \\
2.2 & VPN and tunneling & Infrastructure that routes some or all of a user's traffic through an intermediary connection for the duration of a session, rather than rotating per request as a proxy does. Assign 2.1 separately if proxy/SOCKS services are also explicitly offered. & ``check out your country tweak, download our latest vpn -- new servers, fixed bugs, optimized apk'' & VPN, tunnel, SSH tunnel, port forwarding \\
\multicolumn{5}{@{}l}{\textit{3. Communication Infrastructure}} \\
3.1 & Mail infrastructure & Infrastructure for sending, relaying, or managing outbound email at volume (SMTP access, mail relays, mailer software, or bulk/campaign delivery platforms), regardless of the email's purpose. Do not assign for offers of email/phone lead lists or stolen credentials that do not include mail-sending infrastructure itself. & ``smtp stock on sale: rackspace, bulletproof, sendinblue, aws ses, all regions available'' & SMTP, mailer, mail relay, inboxing \\
\midrule
\multicolumn{5}{@{}l}{\textit{Trust attributes (not mutually exclusive):}} \\
-- & Bulletproof & An explicit marketing claim that a service will keep operating despite abuse complaints, law-enforcement requests, or takedown/suspension attempts. Coded per message, independent of which infrastructure category is advertised. & ``senders/mailers/cpanels/
validator/rdp/smtps -- bulletproof hosting, hitting inbox'' & Bulletproof, no logs, offshore \\
-- & Payment security & A claim or mechanism that reduces transaction risk for buyer or seller, most commonly escrow, but also cryptocurrency payment, digital-wallet options, or automated/instant payment processing. Coded per message, independent of infrastructure category. & ``i accept any crypto -- escrow accepted'' & Escrow, crypto payment, digital wallet \\
-- & Transparency & A claim signaling accountability or reliability of service, such as guaranteed customer support, proof of service or delivery, or warranties. Coded per message, independent of infrastructure category. & ``24/7 customer support -- 1 month guaranteed'' & Customer support, guarantee, proof of delivery, warranty \\
\bottomrule
\end{tabular}
\end{table*}

\subsection{Inter-Annotator Agreement}
\label{app:kappa}
Tab.~\ref{tab:iaa} reports Cohen's kappa by category and attribute, computed before adjudication (Section~\ref{sec:methodology}).
\begin{table}
\caption{Inter-annotator agreement by category and attribute
(250-message taxonomy validation sample).}
\label{tab:iaa}
\small
\begin{tabular}{lrrrrrr}
\toprule
Category/Attribute & Kappa & Acc. & Prec. & Rec. & F1 & Disagr. \\
\midrule
1.1 Hosting services      & 0.744 & 0.872 & 0.798 & 0.934 & 0.861 & 32 \\
1.2 VPS                   & 0.937 & 0.984 & 0.921 & 0.972 & 0.946 & 4  \\
1.3 RDP                   & 0.983 & 0.992 & 0.980 & 1.000 & 0.990 & 2  \\
2.1 Proxy services        & 0.576 & 0.972 & 0.417 & 1.000 & 0.588 & 7  \\
2.2 VPN and tunneling     & 0.823 & 0.936 & 0.773 & 0.981 & 0.864 & 16 \\
3.1 Mail infrastructure   & 0.936 & 0.968 & 0.941 & 1.000 & 0.970 & 8  \\
Bulletproof               & 0.895 & 0.948 & 0.939 & 0.947 & 0.943 & 13 \\
Payment security          & 0.786 & 0.960 & 0.724 & 0.913 & 0.808 & 10 \\
Transparency               & 0.644 & 0.824 & 0.913 & 0.700 & 0.792 & 44 \\
\midrule
Macro avg. & 0.814 & 0.940 & 0.823 & 0.939 & 0.862 & -- \\
Micro avg. & 0.859 & 0.940 & --    & --    & 0.902 & -- \\
\bottomrule
\end{tabular}
\end{table}

\section{Detection Methods Comparison}
\label{app:automation_methods}

This appendix supplements Section~\ref{sec:labeling} with the classification prompt, implementation details and additional evaluation results. The six infrastructure categories were treated as a multi-label task, while Bulletproof, Payment Security and Transparency were treated as separate binary targets.



\subsection{Methodology for Detection Methods}
\label{app:detection_comparison}

\paragraph{Evaluation setup.}
The keyword and TF-IDF approaches used a fixed 75\%-25\% train-test split.
Both were trained only on the training split and evaluated on the same
held-out messages. For the infrastructure-category task, we report
micro-F1, macro-F1, micro-precision, micro-recall,
sample-averaged Jaccard similarity, and exact-match accuracy. Micro-averaged metrics measure aggregate
performance across all labels, whereas macro-F1 gives equal weight to each
label and is therefore more sensitive to performance on infrequent
categories. Exact match requires the complete predicted label set for a
message to equal its ground-truth label set. The three trust attributes
were evaluated independently using precision, recall, F1, and accuracy.

\paragraph{Keyword dictionary.}
Messages were normalized by converting text to lowercase, removing URLs
and unsupported characters, and collapsing repeated whitespace. For each
infrastructure category, the dictionary combined manually selected seed
terms, category-specific expressions from the annotated \texttt{Keywords}
field of the ground-truth messages in the training set, and discriminative
unigrams and bigrams learned exclusively from that training set. Candidate
n-grams were extracted using a maximum vocabulary of 4,000 features. Up to
20 learned terms were retained per label when their normalized occurrence
rate in positive examples was at least 1.5 times that in negative
examples. Stopwords and short non-informative terms were removed. The
resulting terms were converted into label-specific regular expressions,
and every label with a matching expression was assigned. Trust attributes
were processed the same way, using separate dictionaries.

\paragraph{TF-IDF with logistic regression.}
Messages were normalized using the same procedure and represented with
word-level TF-IDF features (unigrams and bigrams, maximum 20,000 features,
\texttt{min\_df=1}, \texttt{max\_df=0.98}, sublinear term-frequency
scaling). The six infrastructure categories were encoded as independent
binary targets and classified using one-vs-rest logistic regression with
balanced class weights, \(C=2.0\), the \texttt{liblinear} solver, a
maximum of 2,000 iterations, and random seed 42. A separate vectorizer and
classifier set was trained for the three trust attributes. A label was
assigned when its predicted probability was at least 0.5.

\paragraph{Prompt-based LLM classification.}
We evaluated three locally deployable, open-weight models --
Mistral-Nemo-12B, DeepSeek-R1-Distill-Qwen-14B, and Qwen2.5-14B -- rather
than a frontier model, since our goal was a reproducible pipeline under
modest compute rather than an upper bound on LLM classification
performance. Each model received one Telegram message together with the
taxonomy definitions, category boundaries, and multi-label decision rules,
and was instructed to return all applicable labels in a fixed JSON object,
or an empty list if none applied. Messages were processed independently,
without information from previously classified messages. The complete
prompt is provided in the repository.

\paragraph{Results.}
The main-text comparison in Tab.~\ref{tab:classifier-comparison-main} shows that TF-IDF achieves higher micro-F1 and precision, whereas the keyword approach achieves higher macro-F1 and recall. This trade-off indicates that stronger aggregate performance does not imply uniformly better performance across categories. The exploratory zero-shot LLM evaluation yielded micro-F1 scores of 0.401 for Mistral-Nemo-12B, 0.309 for DeepSeek-R1-Distill-Qwen-14B and 0.278 for Qwen2.5-14B. These results characterize the evaluated models and prompting setup rather than LLM classification performance in general.

\begin{table}[t]
\caption{LLM performance on 261 human-annotated messages.}
\label{tab:llm}
\centering
{\setlength{\tabcolsep}{1.6pt}
\begin{tabular}{@{}lrrrrrrrr@{}}
\toprule
Model & Mi-P & Mi-R & Mi-F1 & Ma-P & Ma-R & Ma-F1 & Jac. & EM \\
\midrule
Mistral-NeMo & 0.494 & 0.337 & 0.401 & 0.272 & 0.281 & 0.272 & 0.267 & 0.027 \\
DeepSeek-R1  & 0.438 & 0.238 & 0.309 & 0.351 & 0.163 & 0.214 & 0.164 & 0.000 \\
Qwen2.5      & 0.413 & 0.210 & 0.278 & 0.398 & 0.143 & 0.195 & 0.147 & 0.027 \\
\bottomrule
\end{tabular}}
\end{table}
\paragraph{Trust-attribute evaluation.}
We apply the same three-way comparison to the three trust attributes using
the ground truth in Section~\ref{sec:groundtruth}. This is deliberately
lighter-weight than \cite{ricaldi2025trust}'s approach of fine-tuning a
large model (DeepSeek-V3) for trust-signal labeling: repeating that
process at 1.1M-message scale, and again each time the signal set changes
-- as it does here with the addition of Bulletproof -- is costly in both
compute and annotation time, and not sustainable as the taxonomy is
extended. A lightweight classifier is far cheaper to (re-)train as the
taxonomy evolves.

\section{Classifier Validation with Independent Human Annotation}
\label{app:human-validation}

To complement the inter-annotator agreement reported in Tab.~\ref{tab:iaa}, which measures agreement
between coders on a 250-message set, we additionally validated the
trained classifier's predictions against independent human annotation
on 200 further messages not included in the taxonomy validation set,
annotated by a human annotator using the same codebook
(App.~\ref{app:codebook}). Three subsets were
constructed: (i) a domain-transfer sample (n=100) drawn from
communities outside the 11 used for taxonomy development and
training, restricted to messages the classifier predicted as
infrastructure, to assess precision under domain shift; (ii) a
predicted-infrastructure sample (n=50) from the training domain, also
restricted to positive predictions, to estimate in-domain precision;
and (iii) a predicted-non-infrastructure sample (n=50) to assess the frequency of missed positives among sampled predicted-negative messages. The support (n) in
Tab.~\ref{tab:val-domain-transfer} and~\ref{tab:val-predicted-infra}
is the number of positive messages assess the frequency of missed positives among sampled predicted-negative messagesannotated for that category.

\subsection{Domain-Transfer Validation (Precision Under Domain Shift)}

Because every message in this sample was predicted positive by the
classifier, only Precision is an interpretable summary of the overall
Infrastructure label (marked~$\dagger$); Recall and Accuracy are fixed
by construction. Precision on the overall Infrastructure label was
78.0\%. Category-level agreement ranged from fair (Hosting,
$\kappa=0.307$) to substantial (Mail, $\kappa=0.678$; VPS,
$\kappa=0.677$); VPN had no positive examples in this sample and Proxy
had only one.

\begin{table}[h]
\caption{Domain-transfer validation: classifier vs.\ human annotation
by category (n=100).}
\label{tab:val-domain-transfer}
\small
\begin{tabular}{lrrrrr}
\toprule
Category & $\kappa$ & Prec. & Rec. & F1 & Support \\
\midrule
Infrastructure$^\dagger$ & -- & 0.780 & -- & -- & 78 \\
Hosting                  & 0.307 & 0.412 & 0.808 & 0.545 & 26 \\
VPS                      & 0.677 & 0.636 & 0.875 & 0.737 & 16 \\
RDP                      & 0.529 & 0.574 & 0.900 & 0.701 & 30 \\
VPN                      & --    & --    & --    & --    & 0  \\
Proxy                    & 1.000 & 1.000 & 1.000 & 1.000 & 1  \\
Mail                     & 0.678 & 0.803 & 0.983 & 0.884 & 58 \\
\bottomrule
\end{tabular}
\\[2pt]
\footnotesize $\dagger$ All messages in this sample were predicted
positive; only Precision is interpretable. Multi-label (pooled):
Cohen's $\kappa=0.653$, Micro F1$=0.743$, Macro F1$=0.645$, Exact
Match Accuracy$=0.310$.
\end{table}

\subsection{Predicted-Infrastructure Validation (In-Domain Precision)}

The same overall-label caveat applies here. Precision on the overall
Infrastructure label was 84.0\%. Mail ($\kappa=0.831$) and Hosting
($\kappa=0.647$) showed the strongest agreement; RDP was weakest
($\kappa=0.323$), the inverse of its domain-transfer ranking.

\begin{table}[h]
\caption{Predicted-infrastructure validation: classifier vs.\ human
annotation by category (n=50).}
\label{tab:val-predicted-infra}
\small
\begin{tabular}{lrrrrr}
\toprule
Category & $\kappa$ & Prec. & Rec. & F1 & Support \\
\midrule
Infrastructure$^\dagger$ & -- & 0.840 & -- & -- & 42 \\
Hosting                  & 0.647 & 0.724 & 0.955 & 0.824 & 22 \\
VPS                      & 0.638 & 0.500 & 1.000 & 0.667 & 3  \\
RDP                      & 0.323 & 0.316 & 0.857 & 0.462 & 7  \\
VPN                      & 0.000 & 0.000 & 0.000 & 0.000 & 0  \\
Proxy                    & --    & --    & --    & --    & 0  \\
Mail                     & 0.831 & 0.925 & 1.000 & 0.961 & 37 \\
\bottomrule
\end{tabular}
\\[2pt]
\footnotesize $\dagger$ All messages in this sample were predicted
positive; only Precision is interpretable. Multi-label (pooled):
Cohen's $\kappa=0.743$, Micro F1$=0.812$, Macro F1$=0.485$, Exact
Match Accuracy$=0.560$.
\end{table}

\subsection{Missed Infrastructure Among Predicted-Negative Messages}

Since every message in this sample was predicted negative on all six
categories, the informative statistics are the raw miss counts. Of
50 messages the classifier predicted as non-infrastructure, human
annotation identified 5 (10.0\%) containing at least one
infrastructure category. Across the 300 individual category decisions
(50 messages~$\times$~6 categories), 295 (98.3\%, Hamming Accuracy)
matched; 45 of 50 messages (90.0\%, Exact Match Accuracy) matched on
all six categories.

\subsection{Trust-Attribute Validation}
We extended this validation to the three trust attributes, applying the
same logic to a separate sample of 100 unique messages, stratified
across the classifier's predicted-positive pool for each attribute
(Bulletproof, Payment security, Transparency) to ensure adequate
support for Payment security, the rarest of the three, and
independently annotated using the same codebook. Tab. \ref{tab:trust_validation} shows the results of this validation process.

\begin{table}[t]
\centering
\caption{Trust-attribute validation: classifier vs.\ human annotation (n=100).}
\label{tab:trust_validation}
\begin{tabular}{@{}lccccc@{}}
\toprule
Attribute & $\kappa$ & Prec. & Rec. & F1 & Support \\
\midrule
Bulletproof      & 0.762 & 0.769 & 1.000 & 0.870 & 40 \\
Payment security & 0.342 & 0.343 & 0.800 & 0.480 & 15 \\
Transparency     & 0.640 & 0.833 & 0.800 & 0.816 & 50 \\
\bottomrule
\end{tabular}
\vspace{2pt}

{\footnotesize Multi-label (pooled): Cohen's $\kappa$ = 0.615, Micro F1 = 0.767, Macro F1 = 0.722, Exact Match Accuracy = 0.510.}
\end{table}

Pooled agreement was substantial ($\kappa$ = 0.615). Bulletproof and
Transparency showed substantial agreement ($\kappa$ = 0.762 and 0.640);
Payment security showed only fair agreement ($\kappa$ = 0.342), driven
by low precision on the smallest support of the three (n = 15).

\section{Group-Level Infrastructure Prevalence}
\label{app:details_results}

Tab.~\ref{tab:group_level_infrastructure_overview} summarizes infrastructure activity at the Telegram group level. Predicted infrastructure activity was present in 113 of the 167 Telegram groups, corresponding to 67.66\% of all groups. These infrastructure-active groups contain 99.03\% of all messages in the dataset. However, the difference between the average and median number of infrastructure messages per infrastructure-active group indicates that activity is highly unevenly distributed.
\begin{table}[t]
\centering
\footnotesize
\caption{Group-level overview of predicted infrastructure-related messages.}
\label{tab:group_level_infrastructure_overview}
\setlength{\tabcolsep}{4pt}
\renewcommand{\arraystretch}{1.15}
\begin{tabularx}{\columnwidth}{@{}>{\raggedright\arraybackslash}X r@{}}
\toprule
\textbf{Metric} & \textbf{Value} \\
\midrule
Total number of groups & 167 \\
Groups with at least one infrastructure message & 113 (67.66\%) \\
Total number of messages & 1,116,071 \\
Infrastructure messages & 207,244 (18.57\%) \\
Average infrastructure messages per infrastructure-active group & 1,834 \\
Median infrastructure messages per infrastructure-active group & 11 \\
Maximum infrastructure messages in one group & 104,320 \\
Maximum infrastructure share in one group & 58.49\% \\
\bottomrule
\end{tabularx}
\end{table}

Of the 207,244 messages assigned at least one infrastructure category, 69.1\% received multiple labels (mean: 1.95 categories), consistent with bundled advertising. Activity was highly uneven across communities: 82 of the 113 infrastructure-active communities (72.6\%) contained fewer than 50 classified messages, whereas only five contained at least 5,000. Thus, infrastructure advertising was broadly present but concentrated by volume in a small number of communities.

\subsection{Case Study: Vendor Composition of the Highest-Volume
Community}
\label{app:case-study}

The highest-volume community (\S\ref{sec:results-concentration}) accounts for 50.34\% of all
classified infrastructure messages (104,320 of 207,244) and hosts 562
distinct vendors. Its within-community vendor Gini coefficient
(0.809) is lower than the overall vendor-level Gini across all
communities (0.884, \S\ref{sec:results-vendor}), reinforcing that this community's
dominance reflects a broad vendor base rather than one or two sellers
driving its volume: the top~10 vendors jointly account for 13.77\% of
all classified infrastructure messages platform-wide (27.35\% of the
community), and the top~50 for 32.31\% platform-wide (64.18\% of the
community).

Tab.~\ref{tab:case-study-vendors} lists the ten most active vendors
in this community, anonymized as V1--V10. Most advertise Mail
infrastructure alongside Hosting and/or VPS. Two of the five most
active vendors are also active in additional communities, meaning
disruption of this community alone would not remove them from the
ecosystem.

\begin{table}
\caption{Top 10 vendors in the highest-volume community.} \label{tab:case-study-vendors}
\small
\begin{tabular}{lrrr}
\toprule
Vendor & Msgs. in comm. & \% of comm.'s msgs. & Comm. active in \\
\midrule
V1  & 5,451 & 5.23\% & 1 \\
V2  & 3,611 & 3.46\% & 2 \\
V3  & 3,525 & 3.38\% & 1 \\
V4  & 2,830 & 2.71\% & 3 \\
V5  & 2,564 & 2.46\% & 1 \\
V6  & 2,226 & 2.13\% & 1 \\
V7  & 2,218 & 2.13\% & 1 \\
V8  & 2,175 & 2.08\% & 2 \\
V9  & 2,134 & 2.05\% & 1 \\
V10 & 1,799 & 1.72\% & 1 \\
\bottomrule
\end{tabular}
\end{table}

\end{document}